\documentclass[aps,prl,twocolumn,groupedaddress]{revtex4}
\usepackage{graphicx}
\begin{document}

\title{Duality of liquids}
\author{K. Trachenko$^{*1}$}
\author{V. V. Brazhkin$^{2}$}
\address{$^1$ School of Physics and Astronomy, Queen Mary University of London, Mile End Road, London, E1 4NS, UK, email k.trachenko@qmul.ac.uk}
\address{$^2$ Institute for High Pressure Physics, RAS, 142190, Moscow, Russia}

\begin{abstract}
Liquids flow, and in this sense are close to gases. At the same time, interactions in liquids are strong as in solids. The combination of these two properties is believed to be the ultimate obstacle to constructing a general theory of liquids. Here, we adopt a new approach: instead of focusing on the problem of strong interactions, we zero in on the relative contributions of vibrational and diffusional motion. We show that liquid energy and specific heat are given, to a very good approximation, by their vibrational contributions as in solids over almost entire range of relaxation time in which liquids exist as such, and demonstrate that this result is consistent with liquid entropy exceeding solid entropy. Our analysis therefore reveals an interesting {\it duality} of liquids not hitherto known: they are close to solids from the thermodynamic perspective and to flowing gases. We discuss several implications of this result.
\end{abstract}

\maketitle

\section{Introduction}
The development of basic theories of solids involved a number of important discoveries that date back to over 100 years. This was preceded by the development of the theories of gases. These developments form the basis for current understanding of most essential properties of these two basic states of matter \cite{landau}. The third state of matter, the liquid state, remains poorly understood in comparison. A testament to this comes from a surprising fact that even recent textbooks dedicated to liquids do not discuss most basic liquid properties such as specific heat \cite{hansen}. Surprising though it may seem to a scientist outside the area, this fact has long been appreciated by those teaching the subject. In an amusing story about student teaching experience, Granato recalls his persistent fear of a potential student question about liquid specific heat \cite{granato}. Noting that such a question was never asked over many years by a total of 10,000 students, Granato observes that this possibly reflects an important deficiency of our standard teaching method that fails to mention unsolved problems in physics, both in lectures and textbooks.

Liquids flow, and share this property with gases. At the same time, interactions in a liquid are strong, and are similar to those in solids. This presents a fundamental difficulty in calculating liquid energy in general form. Indeed, strong interactions, combined with system-specific form of interactions, imply that the energy is strongly system-dependent, ostensibly precluding the calculation of energy in general form, contrary to solids or gases \cite{landau}.

Strong interactions are successfully treated in solids in the phonon approach, but this approach has long been thought to be inapplicable to liquids where atomic displacements are large. Stated differently, the ``small parameter'' in the theory of solids are atomic displacements, and a harmonic contribution to the energy, the phonon energy, is often a good approximation. The small parameter in gases are weak interatomic interactions. On the other hand, liquids have none of these because interactions are strong and displacements are large. The absence of a small parameter was, in Landau view, the fundamental property of liquids that ultimately precluded the construction of a theory of liquids at the same level existing for solids or gases \cite{landau}.

Here, we propose that reformulating the problem and exploring it in the new formulation provides an important way in. Instead of starting at the level of strong interactions at the atomistic scale, we focus on the atomic trajectories that result from these interactions. This reason is not unrelated to the one stated by Landau above, but operates at a different level.

We have recently proposed \cite{phystoday} that from the point of atomic dynamics, solids and gases are pure states of matter in the sense that dynamics in solids is purely oscillatory and dynamics in gases are purely ballistic and collisional. The dynamics of a liquid, on the other hand, is not pure but mixed: it involves both oscillations and ballistic motions, and the relative contributions of the two types of motion change in response to external parameters, temperature and pressure. Physically, the different behaviors arise because in solids the kinetic energy of particles, $K$, is much smaller than the energy barriers between various potential minima, $U$: $K\ll U$. In gases, it is the other way around: $K\gg U$. In liquids, the mixed nature of dynamics originates because none of these strong inequalities apply.

Notably, the solid-like oscillatory component of liquid dynamics originates from large energy barriers preventing local atomic jumps and diffusion processes. Large energy barriers, in turn, are set by strong interactions - this is how interaction strength, emphasized by Landau, enters our approach.

In our approach to liquid thermodynamics, we therefore focus on atomic trajectories and relative weights of solid-like oscillatory and gas-like diffusional motions. In this approach, we present a proof that liquid energy and heat capacity are given, to a very good approximation, by their vibrational contributions as in solids over almost entire range of relaxation times in which liquids exist as such. We show that this result is consistent with liquid entropy exceeding solid entropy. Our analysis therefore shows that while liquids are close to gases from the point of view of flow, they are also close to solids from the thermodynamic perspective. This reveals a new property of liquids not hitherto known: their {\it duality}.

We note that the possibility of oscillatory motion contributing to liquid heat capacity has been contemplated before. This was done on the basis of empirical observation that experimental specific heat of some monatomic liquids (e.g. liquid metals) around the melting point is close to the Dulong-Petit value \cite{agren,grim,wallace}. Our new result here is the rigorous proof that liquid heat capacity is given by the vibrational motion over almost entire range of liquid relaxation time. This is an important advance in view of the absence of a theory of liquids \cite{granato}.

\section{Results}

\subsection{Energy and specific heat}

Our first step is to calculate the liquid energy by evaluating relative contributions of oscillatory and ballistic diffusional motions. The evaluation can be done using the concept of liquid relaxation time proposed by Frenkel \cite{frenkel}. Atoms or molecules in a liquid are not fixed, but rearrange in space due to thermally activated processes, giving liquid flow. Each flow event is a jump of an atom from its surrounding cage, accompanied by large-scale rearrangement of the cage atoms. We call this process a local relaxation event (LRE). Frenkel introduced liquid relaxation time $\tau$ as the time between LREs at one point in space in a liquid \cite{frenkel}, and showed that $\tau$ is related to liquid viscosity $\eta$ via the Maxwell relationship $\eta=G_{\infty}\tau$, where $G_{\infty}$ is the instantaneous shear modulus.

The concept of $\tau$ has been widely used since to discuss liquid dynamics and its changes with temperature \cite{dyre,angell}. In this picture, liquid dynamics acquires a simple description: a particle spends time $\tau$ oscillating inside the cage before jumping to a nearby quasi-equilibrium site. The range of $\tau$ is bound by two important values. At low temperature, $\tau$ increases until it reaches the value at which the liquid stops flowing at the experimental time scale. This corresponds to $\tau\approx 10^{3}$ s and the liquid-glass transition \cite{dyre,angell}. At high temperature, $\tau$ approaches its minimal value given by Debye vibration period, $\tau_{\rm D}\approx 0.1$ ps, when the time between the jumps becomes comparable to the shortest vibrational period. As a result, $\tau$ varies by about 16 orders of magnitude in which the liquid state of flow can be measured.

There are two contributions to liquid energy, $E_{\rm l}$: $E_{\rm l}=E_{\rm vib}+E_{\rm dif}$, where $E_{\rm vib}$ and $E_{\rm dif}$ are the energies of vibrations and diffusion, respectively. $E_{\rm dif}$ includes the kinetic energy of jumping atoms as well as the energy of their interaction with other atoms during the local jump events. $E_{\rm vib}$ has two contributions predicted by Frenkel \cite{frenkel}: one longitudinal mode and two transverse modes with frequency $\omega>\frac{1}{\tau}$. The prediction was made on the basis of observation that at times shorter than $\tau$, a liquid is essentially a frozen solid with all three vibrational modes, including transverse ones. On the other hand, a liquid flows and yields to shear stress at times longer than $\tau$, and therefore does not support transverse modes at frequency $\omega<\frac{1}{\tau}$.

Since Frenkel's prediction, the ability of liquids to support solid-like modes with wavelengths extending to the shortest distance comparable to interatomic separations has been confirmed experimentally \cite{w1,w2,w3,pilgrim,giord,hoso,water}. Notably, most of this experimental evidence is fairly recent, and has started to come to the fore only when powerful synchrotron radiation sources started to be deployed, some 50--60 years after Frenkel's prediction. This long-lived absence of experimental data about propagating collective excitations in liquids may have contributed to their poor understanding from the theoretical point of view.

We note here that contrary to sometimes expressed views \cite{ruocco,egami}, collective excitations in disordered systems such as glasses and liquids do {\it not} decay or get damped, and are propagating up to the shortest wavelengths. Indeed, disordered systems support non-decaying solutions, eigenstates, of the secular equation involving the force matrix constructed from the disordered structure (this structure is static in a glass whereas in a liquid it is static at times smaller than $\tau$ as discussed by Frenkel). The energy of the disordered system is then equal to the energy of non-decaying eigenstate collective excitations. Harmonic (plane-wave) excitations, including those measured by the experimental probes, naturally decay in disordered systems, yet importantly these are clearly seen in liquids as solid-like quasi-linear dispersion curves up to the shortest wavelengths \cite{w1,w2,w3,pilgrim,giord,hoso,water}. This includes transverse waves with the shortest wavelengths comparable to interatomic separations that are seen in even low-viscous liquids such as liquid Na \cite{giord}, Ga \cite{hoso}, water \cite{water} and so on, consistent with Frenkel's prediction. A detailed discussion of this point is forthcoming. Below, we will make use of the fact that the experimental solid-like quasi-linear dispersion curves in liquids imply that their vibrational density of states can be approximated by the quadratic form $g(\omega)\propto\omega^2$ to the same extent as in solids.

Lets now consider the regime where LREs take place rarely compared to the short period of vibrations:

\begin{equation}
\frac{\tau_{\rm D}}{\tau}\ll 1
\label{3}
\end{equation}

The jump probability for a LRE, $\rho$, is the ratio between the time spent diffusing and vibrating. A LRE lasts on the order of Debye vibration period $\tau_{\rm D}\approx 0.1$ ps. Therefore, $\rho=\frac{\tau_{\rm D}}{\tau}$. In statistical equilibrium, $\rho$ is equal to the ratio of diffusing atoms, $N_{\rm dif}$, and the total number of atoms, $N$. Then, at any given moment of time:

\begin{equation}
\frac{N_{\rm dif}}{N}=\frac{\tau_{\rm D}}{\tau}
\label{1}
\end{equation}

If $E_{\rm dif}$ is the energy associated with diffusing LREs, $E_{\rm dif}\propto N_{\rm dif}$. Together with $E_{\rm tot}\propto N$, Eq. ({\ref{1}) gives

\begin{equation}
\frac{E_{\rm dif}}{E_{\rm tot}}=\frac{\tau_{\rm D}}{\tau}
\label{2}
\end{equation}

Eq. (\ref{2}) implies that under condition (\ref{3}), the contribution of $E_{\rm dif}$ to the total energy at any moment of time is negligible. We note that Eq. (\ref{2}) corresponds to the instantaneous value of $E_{\rm dif}$ which, from the physical point of view, is given by the smallest time scale of the system, $\tau_{\rm D}$. During time $\tau_{\rm D}$, the system is not in equilibrium. The equilibrium state is reached when the observation time exceeds system relaxation time, $\tau$. After time $\tau$, all LREs in the system relax. Therefore, we need to calculate $E_{\rm dif}$ that is averaged over time $\tau$.

Let us divide time $\tau$ into $m$ time periods of duration $\tau_{\rm D}$ each, so that $m=\frac{\tau}{\tau_{\rm D}}$. Then, $E_{\rm dif}$, averaged over time $\tau$, $E^{\rm av}_{\rm dif}$, is

\begin{equation}
E^{\rm av}_{\rm dif}=\frac{E^1_{\rm dif}+E^2_{\rm dif}+...+E^m_{\rm dif}}{m}
\label{ave}
\end{equation}

\noindent where $E^i_{\rm dif}$ are instantaneous values of $E_{\rm dif}$ featured in Eq. (\ref{2}). $\frac{E^{\rm av}_{\rm dif}}{E_{\rm tot}}$ is

\begin{equation}
\frac{E^{\rm av}_{\rm dif}}{E_{\rm tot}}=\frac{E^1_{\rm dif}+E^2_{\rm dif}+...+E^m_{\rm dif}}{{E_{\rm tot}}\cdot m}
\label{ave01}
\end{equation}

Each of the terms $\frac{E^i_{\rm dif}}{E_{\rm tot}}$ in Eq. (\ref{ave01}) is equal to $\frac{\tau_{\rm D}}{\tau}$, according to Eq. (\ref{2}). There are $m$ terms in the sum in Eq. (\ref{ave01}). Therefore,

\begin{equation}
\frac{E^{\rm av}_{\rm dif}}{E_{\rm tot}}=\frac{\tau_{\rm D}}{\tau}
\label{ave1}
\end{equation}

We therefore find that under the condition (\ref{3}), the ratio of the average energy of diffusion motion to the total energy is negligibly small, as in the instantaneous case. Consequently, the energy of the liquid under the condition (\ref{3}) is, to a very good approximation, given by the remaining vibrational part:

\begin{equation}
E_{\rm l}=E_{\rm l}^{\rm vib}
\label{energy}
\end{equation}

This, in turn, implies that the liquid constant-volume specific heat, $c_{\rm {v,l}}=\frac{1}{N}\frac{{\rm d}E_{\rm l}}{{\rm d}T}$ (here, $N$ is the number of particles), is entirely vibrational in the regime Eq. (\ref{3}):

\begin{equation}
c_{\rm {v,l}}=c_{\rm {v,l}}^{\rm vib}
\label{cv}
\end{equation}

The vibrational energy and specific heat of liquids in the regime (\ref{3}) can be easily ascertained. When the regime (\ref{3}) is operative, $E_{\rm l}^{\rm vib}$ to a very good approximation is $E_{\rm l}^{\rm vib}=3NT$ (here and below, $k_{\rm B}=1$). Indeed, a solid supports one longitudinal mode and two transverse waves in the range $0<\omega<\frac{1}{\tau_{\rm D}}$. The ability of liquids to support shear modes with frequency $\omega>\frac{1}{\tau}$, combined with $\tau\gg\tau_{\rm D}$ in Eq. (1), implies that a liquid supports most of the shear modes present in a solid. Furthermore and importantly, it is only the high-frequency shear modes that make a significant contribution to the liquid vibrational energy, because the density of states of liquid quasi-harmonic modes is approximately proportional to $\omega^2$ as discussed above. Hence in the regime (\ref{3}), $E_{\rm l}^{\rm vib}=3NT$ to a very good approximation, as in a solid.

We now consider Eqs. (\ref{energy},\ref{cv}) in harmonic and anharmonic cases. In the harmonic case, Eqs. (\ref{energy},\ref{cv}) give the energy and specific heat of a liquid as $3NT$ and $3$, respectively, i.e. the same as in a harmonic solid:

\begin{equation}
E_{\rm l}^{\rm h}=E_{\rm s}^{\rm h}=3NT
\label{henergy}
\end{equation}

\begin{equation}
c_{\rm {v,l}}^{\rm h}=c_{\rm {v,s}}^{\rm h}=3
\label{hcv}
\end{equation}

\noindent where ${\rm s}$ corresponds to the solid and ${\rm h}$ to the harmonic case.

In the anharmonic case, Eqs. (\ref{energy}) and Eqs. (\ref{cv}) still hold, but the equality of liquid and solid energies and specific heats holds only approximately because anharmonicity affects the vibrational energy and $c_{\rm v}$ of a liquid and a solid in a different way. In particular, $c_{\rm v}$ is modified by the intrinsic anharmonicity related to softening of vibrational frequencies at constant volume, and can be approximately written as $c_{\rm v}=3(1+\alpha T)$, where $\alpha$ is the coefficient of thermal expansion \cite{tg,andr}. $\alpha$ is generally larger in liquids compared to solids, resulting in larger liquid $c_{\rm v}$ compared to solid $c_{\rm v}$, albeit the term $\alpha T$ is usually small compared to 1.

The primary evidence supporting Eqs. (\ref{cv},\ref{hcv}) and our theory comes from experimental specific heat of liquids. Early measurements were done for liquid metals, and indicated that their specific heat is very close to 3, the Dulong-Petit value \cite{agren,grim,wallace}. This takes place close the melting point where Eq. (1) applies and which, according to our theory, gives Eqs. (\ref{cv},\ref{hcv}) and $c_{\rm {v,l}}\approx 3$. As experimental techniques advanced and gave access to high pressure and temperature, specific heats of many noble, molecular and network liquids were measured in detail in a wide range of parameters including in the supercritical region \cite{nist}. Similarly to liquid metals, the experimental $c_{\rm v}$ of these liquids was found to be very close to 3 at low temperature where Eq. (1) applies (see Ref. \cite{scirep} for a compilation of the NIST and other data of $c_{\rm v}$ for 21 liquids of different types). This universal behavior provides strong support to our theory.

We note that on temperature increase when condition $\tau$ starts to approach $\tau_{\rm D}$ and Eq. (\ref{3}) no longer applies, experimental $c_{\rm {v,l}}$ starts to decrease from about 3 at the melting point to 2 at high temperature \cite{grim,wallace,nist}. We have recently provided a quantitative description of this effect on the basis of the phonon theory of liquid thermodynamics for many different liquids \cite{prb,scirep}. In this theory, the reduction of heat capacity is due to the progressive loss of transverse modes with frequency $\omega>\frac{1}{\tau}$. On further temperature increase when $c_{\rm v}$ decreases from 2 to its ideal-gas value of $\frac{3}{2}$, another mechanism kicks in: the disappearance of the remaining longitudinal mode with the wavelength smaller than the mean-free path of particles \cite{bolm}. The two mechanisms naturally give a crossover of $c_{\rm v}$ at $c_{\rm v}=2$ that we recently discovered \cite{bolm}, the crossover that corresponds to the Frenkel line where many liquid properties change \cite{phystoday,pre}.

\subsection{Entropy}

Interestingly, although Eq. (\ref{ave1}), combined with Eq. (\ref{3}), implies that the energy and $c_{\rm v}$ of a liquid are entirely vibrational as in a solid, this does {\it not} apply to entropy: the diffusional component to entropy is substantial, and can not be neglected (here and below we imply the equilibrium state, the condition for which $t\gg\tau$, where $t$ is observation time).

Indeed, if $Z_{\rm vib}$ and $Z_{\rm dif}$ are the contributions to the partition sum from vibrations and diffusion, respectively, the total partition sum of the liquid is $Z=Z_{\rm vib}\cdot Z_{\rm dif}$. Then, the liquid energy is $E=T^2\frac{\rm d}{{\rm d}T}\left(\ln(Z_{\rm vib}\cdot Z_{\rm dif})\right)=T^2\frac{\rm d}{{\rm d}T}\ln Z_{\rm vib}+T^2\frac{\rm d}{{\rm d}T}\ln Z_{\rm dif}=E_{\rm vib}+E_{\rm dif}$ (here and below, the derivatives are taken at constant volume). Next, $\frac{E^{\rm av}_{\rm dif}}{E_{\rm tot}}\ll 1$ from Eq. (\ref{ave1}) also implies $\frac{E_{\rm dif}}{E_{\rm vib}}\ll 1$, where, for brevity, we dropped the subscript referring to the average. Therefore, the smallness of diffusional energy, $\frac{E_{\rm dif}}{E_{\rm vib}}\ll 1$, gives

\begin{equation}
\frac{\frac{\rm d}{{\rm d}T}\ln Z_{\rm dif}}{\frac{\rm d}{{\rm d}T}\ln Z_{\rm vib}}\ll 1
\label{ent1}
\end{equation}

The liquid entropy, $S=\frac{\rm d}{{\rm d}T}\left(T\ln(Z_{\rm vib}\cdot Z_{\rm dif})\right)$, is:

\begin{equation}
S=T\frac{\rm d}{{\rm d}T}\ln Z_{\rm vib}+\ln Z_{\rm vib}+T\frac{\rm d}{{\rm d}T}\ln Z_{\rm dif}+\ln Z_{\rm dif}
\label{ent2}
\end{equation}

The condition (\ref{ent1}) implies that the third term in Eq. ({\ref{ent2}) is much smaller than the first one, and can be neglected, giving

\begin{equation}
S=T\frac{\rm d}{{\rm d}T}\ln Z_{\rm vib}+\ln Z_{\rm vib}+\ln Z_{\rm dif}
\label{ent3}
\end{equation}

Eq. (\ref{ent3}) implies that the smallness of $E_{\rm dif}$, expressed by Eq. (\ref{ent1}), does {\it not} lead to the disappearance of all entropy terms that depend on diffusion because the term $\ln Z_{\rm dif}$ remains. This term is responsible for the excess entropy of liquid over the solid. On the other hand, the smallness of $E_{\rm dif}$ {\it does} lead to the disappearance of terms depending on $Z_{\rm dif}$ in the specific heat. Indeed, $c_{\rm {v,l}}=T\frac{{\rm d}S}{{\rm d}T}$ (here, $S$ refers to entropy per atom or molecule), and from Eq. (\ref{ent3}), we find:

\begin{equation}
c_{\rm {v,l}}=T\frac{\rm d}{{\rm d}T}\left(T\frac{\rm d}{{\rm d}T}\ln Z_{\rm vib}\right)+T\frac{\rm d}{{\rm d}T}\ln Z_{\rm vib}+T\frac{\rm d}{{\rm d}T}\ln Z_{\rm dif}
\label{ent4}
\end{equation}

Using Eq. (\ref{ent1}) once again, we observe that the third term in Eq. (\ref{ent4}) is small compared to the second term, and can be neglected, giving

\begin{equation}
c_{{\rm v,l}}=T\frac{\rm d}{{\rm d}T}\left(T\frac{\rm d}{{\rm d}T}\ln Z_{\rm vib}\right)+T\frac{\rm d}{{\rm d}T}\ln Z_{\rm vib}
\label{ent5}
\end{equation}
\noindent

As a result, $c_{\rm v}$ does not depend on $Z_{\rm dif}$, and is given by the vibrational terms that depend on $Z_{\rm vib}$ only. As expected, Eq. (\ref{ent5}) is consistent with Eq. (\ref{cv}).

Physically, the inequality of liquid and solid entropies, $S_{\rm l}\ne S_{\rm s}$, is related to the fact that the entropy measures the total phase space available to the system, which is larger in the liquid due to the diffusional component present in Eq. (\ref{ent3}). The diffusional component, $\ln Z_{\rm dif}$, although large, is slowly varying with temperature according to Eq. (\ref{ent1}), resulting in a small contribution to $c_{\rm v}$ (see Eqs. (\ref{ent4}) and (\ref{ent5})) and giving $c_{\rm {v,l}}=c_{\rm {v,s}}$. On the other hand, the energy corresponds to the instantaneous state of the system (or averaged over $\tau$), and is not related to exploring the phase space. Consequently, $E_{\rm l}=E_{\rm vib}$, yielding Eq. (\ref{ent1}) and the smallness of diffusional contribution to $c_{\rm v}$ despite $S_{\rm l}\ne S_{\rm s}$.

\section{Discussion}

On the basis of results in the previous chapter, we conclude that two important properties of a liquid, energy and specific heat, are essentially vibrational, as they are in a solid, provided $\frac{\tau_{\rm D}}{\tau}\ll 1$. For practical purposes, this takes place for $\tau\gtrsim 10\tau_{\rm D}$. Perhaps not widely recognized, the condition $\tau\approx 10\tau_{\rm D}$ holds even for low-viscous liquids such a liquid monatomic metals (Hg, Na, Rb and so on) and noble liquids such as Ar near their melting points \cite{nist,prb,scirep}, let alone for more viscous liquids such as room-temperature olive or motor oil, honey and so on.

Notably, the condition $\tau\gtrsim 10\tau_{\rm D}$ corresponds to almost the entire range of $\tau$ at which liquids exist. This fact was not fully appreciated in earlier theoretical work on liquids. Indeed, on lowering the temperature, $\tau$ increases from its smallest limiting value of $\tau=\tau_{\rm D}\approx 0.1$ ps to $\tau\approx 10^{3}$ s where, by definition, a liquid forms a glass at the glass transition temperature. Here, $\tau$ changes by 16 orders of magnitude. Consequently, the condition $\frac{\tau_{\rm D}}{\tau}\ll 1$, Eq. (\ref{3}), or $\tau\gtrsim 10\tau_{\rm D}$, applies in the range $10^3-10^{-12}$ s, spanning 15 orders of magnitude of $\tau$. This constitutes almost entire range of $\tau$ where liquids exist as such.

That important elements of liquid thermodynamics can be understood on the basis of thermodynamics of solids is a result hitherto not anticipated. Indeed, starting from earlier proposals \cite{hansen,landau,frenkel}, existing theories approach liquids as interacting gases and consequently attempt to calculate liquid energy as an integral of the product of interatomic interactions and correlation functions (see below). The interactions and correlations are often complex and are not generally known, except in the simplest liquids such as Ar. On the other hand, our theory circumvents this problem because the problem is reduced to calculating the vibrational energy only as in solids.

We now make several further observations regarding the implications of our theory. First, our approach provides general insights into the problem of liquid-glass transition, and implies that the theory of solids is a good starting point for discussing important aspects of liquid thermodynamics. This is a useful result in the area where the calculation of thermodynamic properties of viscous liquids approaching glass transition has been a long-standing problem \cite{dyre,angell}.

Second, we can revisit the long-standing and puzzling question of why Phillips constraint theory of glasses \cite{phil} works. Proposed over 30 years ago, the theory is based on the idea that a short-ranged interatomic bond in a glass can be viewed as a rigid mechanical constraint. Equating the number of constraints and degrees of freedom gives the average coordination number $\langle r\rangle=$2.4 at which, the theory proposes, the glass-forming ability is optimized. The constraint theory has since been used to explain other properties of glass transition \cite{philmag}. Importantly, the constraint theory derives its predictions from counting the bonds (constraints) and $\langle r\rangle$ in the solid glass, but subsequently uses $\langle r\rangle$ to predict the properties measured in the liquid state above the liquid-glass transition. This is truly surprising: indeed, in the liquid state above $T_g$, bonds are not intact because atoms rearrange on the experimental time scale, and therefore can no longer be viewed as rigid constraints. This poses an intriguing question of why the constraint theory works.

Our approach to liquids readily answers this question on general grounds: the measured properties operate in the regime where Eq. (\ref{3}) applies. Indeed, glass-forming melts such as silicates, chalcogenides and others are measured at temperatures where they are viscous enough to satisfy Eq. (\ref{3}) \cite{philmag}. Consequently, our theory predicts that if a measured property is related to system energy or specific heat, this property in the liquid state is, to a very good approximation, equal to that in the solid glass, and can therefore be predicted from solid-state properties.

Finally, our approach relates to previous theories of liquids in another interesting way. When interactions and correlations are pair-wise only as in simple noble liquids such as liquid Ar, previous theories attempt to calculate $c_{\rm v,l}$ as

\begin{equation}
c_{\rm v,l}=\frac{3}{2}+\frac{d}{d T}\int g(r) U(r) dV
\label{int}
\end{equation}

\noindent where $g(r)$ is the normalized pair correlation function and $U(r)$ is the interaction potential.

There has been an extensive work on calculating $g(r)$ and related probability functions in liquids, involving Gaussian and other approximations (see, e.g. Ref. \cite{chan}). Using Eq. (\ref{int}) and our solid-like result for $c_{\rm v,l}$, $c_{\rm v,l}=c_{\rm {v,l}}^{\rm vib}$ ($c_{\rm v,l}=3$ in the harmonic case) gives an important way to validate these approximations. In other words, our result opens a new avenue to elucidate the {\it structure} of liquids, a developing area with long history \cite{hansen}.

In summary, we observe that liquids flow, making them remarkably distinct from solids and close to gases. However, from the point of view of thermodynamics, liquid energy and specific heat are, to a very good approximation, equal to those in solids for relaxation times spanning 15 orders of magnitude, or almost entire range in which the liquids exist as such. In this sense, liquids show an interesting duality of physical properties not hitherto anticipated.

\section{Acknowledgements}

K. Trachenko thanks EPSRC for financial support.

\section{Author contributions}

K. T. and V. V. Brazhkin have contributed equally to this work.

\section{Additional information}

Competing financial interests: The authors declare no competing financial interests.

\end{document}